\documentclass[a4paper,12pt]{article}

\usepackage{a4}
\usepackage{comment}
\usepackage{graphics}
\usepackage{graphicx}
\usepackage{latexsym}
\usepackage{amsfonts}
\usepackage{amssymb}
\usepackage{amsmath}
\usepackage{slashed}
\usepackage{feynmp}
\usepackage{hyperref}
\usepackage{url}
\usepackage{color}
\usepackage{cancel}
\usepackage{multirow,array}
\usepackage{dcolumn}
\usepackage{bm}
\usepackage{mathtools}
\usepackage[linesnumbered,ruled,vlined]{algorithm2e}
\SetAlFnt{\footnotesize}
\SetAlCapFnt{\small}
\SetAlCapNameFnt{\small}
\usepackage{authblk}
\usepackage{etoolbox}
\usepackage{float}
\usepackage{array}
\usepackage{longtable}
\usepackage{pdflscape}
\usepackage{booktabs}
\usepackage{multirow}
\usepackage{mathtools} 

\newcommand{\sus}{\scalebox{1}[1]{\_}}

\title{
  Improved sampling bounds and scalable partitioning for quantum circuit cutting beyond bipartitions.
}

\author[1]{Junya Nakamura\thanks{junya.nakamura@pwc.com}}
\affil[1]{Technology Laboratory, PwC Consulting LLC, Tokyo, Japan}
\author[2]{Takahiko Satoh}
\affil[2]{Faculty of Science and Technology, Keio University, Japan}
\author[1]{Shinichiro Sanji}

\begin{document}

\maketitle

\begin{abstract}
We propose a new method for identifying cutting locations for quantum circuit cutting, with a primary focus on partitioning circuits into three or more parts.
Under the assumption that the classical postprocessing function is decomposable, we derive a new upper bound on the sampling overhead resulting from both time-like and space-like cuts.
We show that this bound improves upon the previously known bound by orders of magnitude in cases of three or more partitions.
Based on this bound, we formulate an objective function, $L_Q^{}$, and present a method to determine cutting locations that minimize it.
Our method is shown to outperform a previous approach in terms of computation time.
Moreover, the quality of the obtained partitioning is found to be comparable to or better than that of the baseline in all but a few cases, as measured by $L_Q^{}$.
These results are obtained by identifying cutting locations in a number of benchmark circuits  of the size and type expected in quantum computations that outperform classical computers.
\end{abstract}

\newpage

\section{Introduction}
Quantum circuit cutting has been proposed as an approach to perform quantum computations that require more qubits than available on single quantum devices~\cite{Peng_2020, Mitarai_2021}.
The basic idea is to partition a large quantum circuit into subcircuits that are smaller in terms of the number of qubits and gates, sample them independently on quantum devices, and classically postprocess the outcomes to reconstruct the output of the original circuit.
Quantum circuits can be partitioned using either {\it time-like} cuts, which replace identity channels with measurement and initialization operations~\cite{Peng_2020}, or {\it space-like} cuts, which decompose two-qubit gates into single-qubit operations~\cite{Mitarai_2021}.
The validity of these approaches has been confirmed through several experiments using real quantum devices~\cite{Ayral_2020,Ayral_2021, ying2023experimental, Yamamoto_2023, bechtold2023investigating,Carrera_Vazquez_2024, singh2024experimental, PhysRevA.110.062620}.

While quantum circuit cutting offers the attractive advantage of executing only small subcircuits, it comes with considerable sampling overhead, which appears as the number of circuit runs (often called shots) that is required to achieve a desired accuracy of the final output.
A number of methods have been proposed to reduce the sampling overhead by utilizing tools such as ancilla qubits and real-time classical communication in refs.~\cite{Peng_2020, Lowe:2022lom, harada2024doubly, pednault2023alternativeapproachoptimalwire, brenner2023optimalwirecuttingclassical, Harrow_2025, bechtold2023circuitcuttingnonmaximallyentangled, Bechtold_2024, Bechtold_2025} for time-like cutting and in refs.~\cite{mitarai2021overhead, Piveteau_2024, Ufrecht_2023, ufrecht2024optimal, Harrow_2025,schmitt2025cutting,schumann2025bridgingwiregatecutting} for space-like cutting. In particular, the recent method~\cite{ufrecht2024optimal} has demonstrated that we can reach the optimal quantum overhead in joint gate-cutting without real-time classical communication and that in parallel gate-cutting without both ancilla qubits and real-time classical communication.

In order to effectively utilize circuit cutting in practical applications, not only cutting algorithm but preprocessing and postprocessing algorithms and software can be also important~\cite{tang2021cutqc, Perlin_2021, tang2022scaleqc, PhysRevA.108.022615, Gentinetta_2024, uchehara2022rotationinspiredcircuitcutoptimization, brandhofer2023optimal, chatterjee2022qurzon, lian2023fast, bhoumik2023distributed,beisel2023quantme4vqa,10196555,li2024efficientcircuitwirecutting, Pawar_2024,Seitz_2024, kan2024scalable, Ren_2024, dou2025larqucutnewcuttingmapping,majumdar2022errormitigatedquantumcircuit, tornow2024scalingquantumcomputationsgate, tornow2024quantumclassicalcomputingtensornetworks, Basu_2022, Basu_2024}.
In particular, the identification of cutting locations as a preprocessing algorithm affects not only the width and depth of the resulting subcircuits, but also the sampling overheads, therefore it can significantly impact the execution time of the subsequent computations and the fidelity~\cite{tang2021cutqc, tang2022scaleqc, brandhofer2023optimal, Pawar_2024, kan2024scalable, Ren_2024, Basu_2024}.

In this paper, we study partitioning of large quantum circuits into three or more parts, under the assumption that the classical postprocessing function is decomposable.
First, under this assumption, we derive a new upper bound on the sampling overhead. We show that, in cases of three or more partitions, this bound can improve upon the previously known bound by orders of magnitude.
The previous studies on sampling overhead reduction (mentioned above) have focused solely on bipartitions, with the exception of ref.~\cite{Peng_2020}.
While ref.~\cite{Peng_2020} derives an upper bound on the sampling overhead for the cases of more than two partitions using time-like cuts, our work addresses combinations of time-like and space-like cuts, and also  achieves a substantial improvement in the upper bound.
Second, based on the above sampling overhead, we define a new objective function $L_Q^{}$, which is considered to appropriately capture the sampling overhead in solutions of three or more partitions, and propose a method to identify cut locations that minimize it.
Since the previous methods~\cite{tang2021cutqc, tang2022scaleqc, brandhofer2023optimal, Pawar_2024, kan2024scalable, Ren_2024, qiskit-addon-cutting} formulate the objective function based on the total number of cuts, they may fail to yield optimal solutions in terms of sampling overhead.
To validate our method, we use a number of quantum circuits from the large-scale category of  QASMBench~\cite{li2023qasmbench}, and adopt Qiskit-addon-cutting (QAC)~\cite{qiskit-addon-cutting} as the baseline.
We observe that our method significantly outperforms QAC in terms of computation time. The quality of the solutions with respect to $L_Q^{}$ is found to be comparable to or better than QAC in most of the circuits.

This paper is organized as follows.
In sec.~\ref{sec:overhead}, we provide the new upper bound on the sampling overhead and validate it in numerical simulations.
In sec.~\ref{sec:method}, we define $L_Q^{}$, and describe our partitioning method and its implementation.
In sec.~\ref{sec:qasmbench}, we apply our method to the QASMBench and validate it by comparing with QAC. We conclude in sec.~\ref{sec:conclusion}.

\section{Sampling overhead for three or more partitions}\label{sec:overhead}
In this section we give a formula for sampling overhead applicable to cases with three or more partitions. This is the basis for the objective function used in our partitioning method.

\begin{table}
  \begin{center}
  \begin{tabular}{ccc}
    Type & $\kappa$ & $\tau$ \\
    \hline
    Time-like cut & 4~\cite{Peng_2020} & 2 \\
    Space-like cut (CX, CZ) & 3~\cite{Mitarai_2021} & 1.5 \\
  \end{tabular}
\end{center}
  \caption{Typical decomposition approaches and their overhead factors, $\kappa$ (eq.~\ref{eq:kappa})and $\tau$ (eq.~\ref{eq:tau}).}
  \label{tbl:kappa_tau}
\end{table}

Assume that we start with an initial state $\rho$, apply three unitary operations $U, V, W$, measure all $n$ qubits in the computational basis, and at the end apply  a classical postprocessing function $f: \{0, 1\}^n_{} \to [-1, 1]$ to the measured bitstring $s$. Our goal is to compute the expectation value of an observable $O$, which is related to $f(s)$ by introducing the projector, $P_s^{}$, on the state corresponding to a bitstring $s$ as
\begin{align}
  O = \sum_s f(s) P_s^{}. \label{eq:observable}
\end{align}
The expectation value of $O$ can be computed from
\begin{align}
  \langle O \rangle & = \mathrm{tr}\bigl(O \mathcal{W} \circ \mathcal{V} \circ \mathcal{U}(\rho)\bigr) \\
  & = \sum_s f(s) \mathrm{tr}\bigl(P_s^{} \mathcal{W} \circ \mathcal{V} \circ \mathcal{U}(\rho)\bigr),\label{eq:exp_value}
\end{align}
where $\mathcal{U}, \mathcal{V}$ and $\mathcal{W}$ are unitary channels corresponding to $U, V$ and $W$, respectively. Time-like and space-like cutting may be realized by decompositions of the unitary channel $\mathcal{V}$,
\begin{align}
  \mathcal{V} = \prod_{j=1}^{K} \biggl( \sum_{i_j^{}} a_j^{}(i_j^{}) \mathcal{F}_j^{}(i_j^{}) \biggr).\label{eq:decompositions}
\end{align}
Here, $\mathcal{F}_j^{}(i_j^{})$ are local unitary channels or measurement operations in the case of space-like cutting, and are measurement operations and qubit initialize channels in the case of time-like cutting. $a_j^{}(i_j^{})$ are real coefficients, and $K$ denotes the number of cutting. By substituting eq.~\ref{eq:decompositions} into eq.~\ref{eq:exp_value}, we may obtain partitioned subcircuits.
Note, however, that the substitution  does not always mean that the circuit is being partitioned~\cite{Mitarai_2021, Yamamoto_2023}.
Let us define
\begin{align}
  \kappa_j^{} \coloneq \sum_{i_j^{}} |a_j^{}(i_j^{})|. \label{eq:kappa}
\end{align}
 From the Hoeffding's inequality, the estimated expectation value $\langle \overline{O} \rangle$ will be within $\epsilon$ of the true value $\langle O \rangle$ with probability at least $1-\delta$, namely
\begin{align}
  \mathrm{Pr} \left[ | \langle \overline{O} \rangle - \langle O \rangle | \le \epsilon \right] \geq 1 - \delta,
\end{align}
if the number of sampling $N$ satisfies~\cite{Peng_2020,Mitarai_2021}
\begin{align}
  N \ge 2 \frac{\bigl(\prod_{j=1}^K \kappa_j^{}\bigr)^2_{}}{\epsilon^2_{}} \ln{\frac{2}{\delta}}.\label{eq:overhead_K}
\end{align}
Since $\bigl(\prod_{j=1}^K \kappa_j^{}\bigr)^2_{}=1$ without cutting, the sampling overhead scales as $\mathcal{O}(\bigl(\prod_{j=1}^K \kappa_j^{}\bigr)^2_{})$.

\begin{figure}
\begin{center}
\includegraphics[width=5cm]{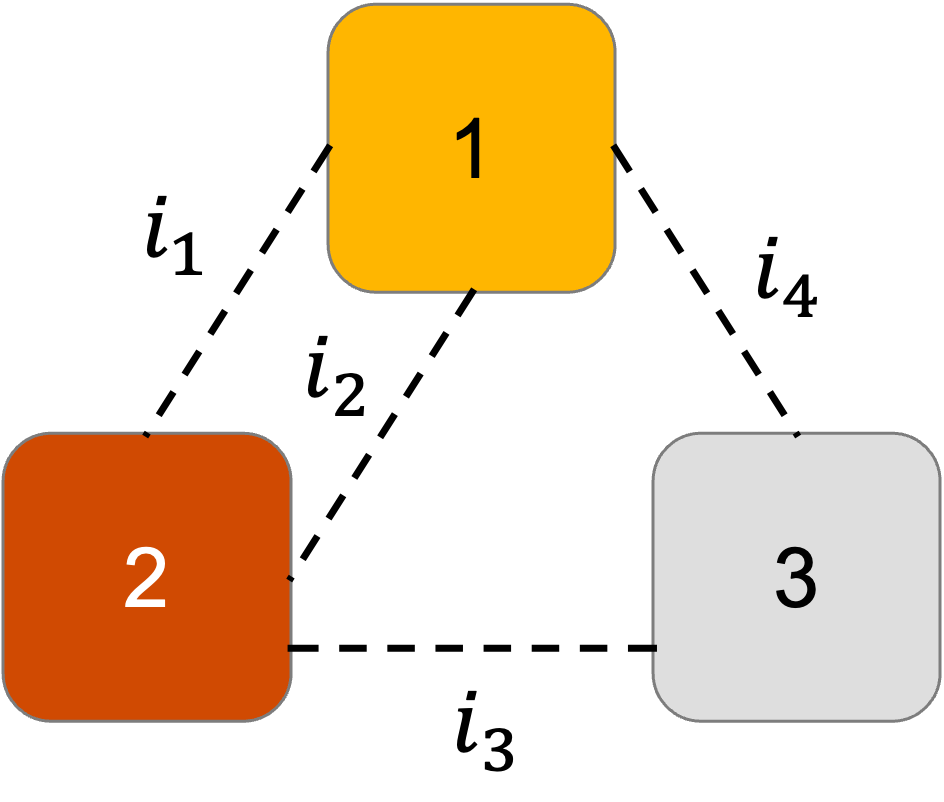}
\end{center}
\caption{A schematic picture showing a circuit divided into three partitions ($R=3$) by four cuts ($K=4$).}
\label{fig:3partition}
\end{figure}

Now we consider the case that the postprocessing function is decomposable into $R$ functions,
\begin{align}
 s =  \prod_{c=1}^{R} s_c^{},\ \  f(s) = \prod_{c=1}^{R}f_c^{}(s_c^{}).\label{eq:f_decompose}
\end{align}
Typical examples include Pauli observables in variational quantum algorithm and estimating probability densities of specific output bitstrings~\cite{Peng_2020}. When considering applications for noisy intermediate scale quantum (NISQ) devices, it is likely that most cases fall into this class, making this assumption reasonable.
As shown in Appendix~\ref{sec:appendix}, the standard deviation of the expectation value $\langle O \rangle$ can be bounded by $\epsilon$, if the total number of sampling $N$ satisfies
\begin{subequations}\label{eq:N_tot}
\begin{align}
  N & \ge \sum_{c=1}^R N_c^{}, \\
  N_c^{} & \ge \frac{R}{\epsilon^2_{}} \Bigl( \prod_{j \in E_c^{}} \kappa_j^{} \Bigr)^2_{}   \prod_{k \in D_c^{}} \tau_k^{}, \label{eq:N_r}
\end{align}
\end{subequations}
where $N_c^{}$ denotes the number of sampling for partition $c$, $E_c^{}$ denotes a set of cuts attached to the partition $c$, $D_c^{}$ denotes a set of cuts not attached to it, and
\begin{align}
  \tau_j^{} \coloneq \sum_{i_j^{}} (a_j^{}(i_j^{}))^2_{}. \label{eq:tau}
\end{align}
Typical decomposition operations and their overhead factors are summarized in Table~\ref{tbl:kappa_tau}.
Eq.~\ref{eq:N_r} indicates that each partition incurs a different overhead and is largely dependent on the cuts directly attached to it. It also depends on cuts not attached to it, but to a lesser extent.
We note that, in the case of a bipartition with only one space-like cutting, our result in eq.~\ref{eq:N_tot} coincides with the formula shown in ref.~\cite{Ufrecht_2023}. Therefore, our result can be regarded as an extension of the study in ref.~\cite{Ufrecht_2023}.

Both eq.~\ref{eq:N_tot} and eq.~\ref{eq:overhead_K}, which is the previously established result, can give the upper bound on the sampling overhead.
To see the difference between these two bounds, let us consider a concrete example of $R=3$ and $K=4$ as illustrated in Figure~\ref{fig:3partition}.
Assuming all the cuts are time-like cuts,  eq.~\ref{eq:N_tot} gives
\begin{align}
 N_{\mathrm{tot}}^{} = N_1^{} + N_2^{} + N_3^{} \ge \frac{52224}{\epsilon^2_{}}.\label{eq:Ntot_our}
\end{align}
This value can be compared with the estimate obtained from eq.~\ref{eq:overhead_K} by setting $\delta=1/3$. Noting that the sampling number derived from eq.~\ref{eq:overhead_K} is required for each partition~\cite{Peng_2020}, we find
\begin{align}
 N_{\mathrm{tot}}^{} \ge \frac{704548}{\epsilon^2_{}}.
\end{align}
These results clearly show that our result eq.~\ref{eq:N_tot} provides a more improved upper bound.

\begin{figure}
\begin{center}
\includegraphics[width=16cm]{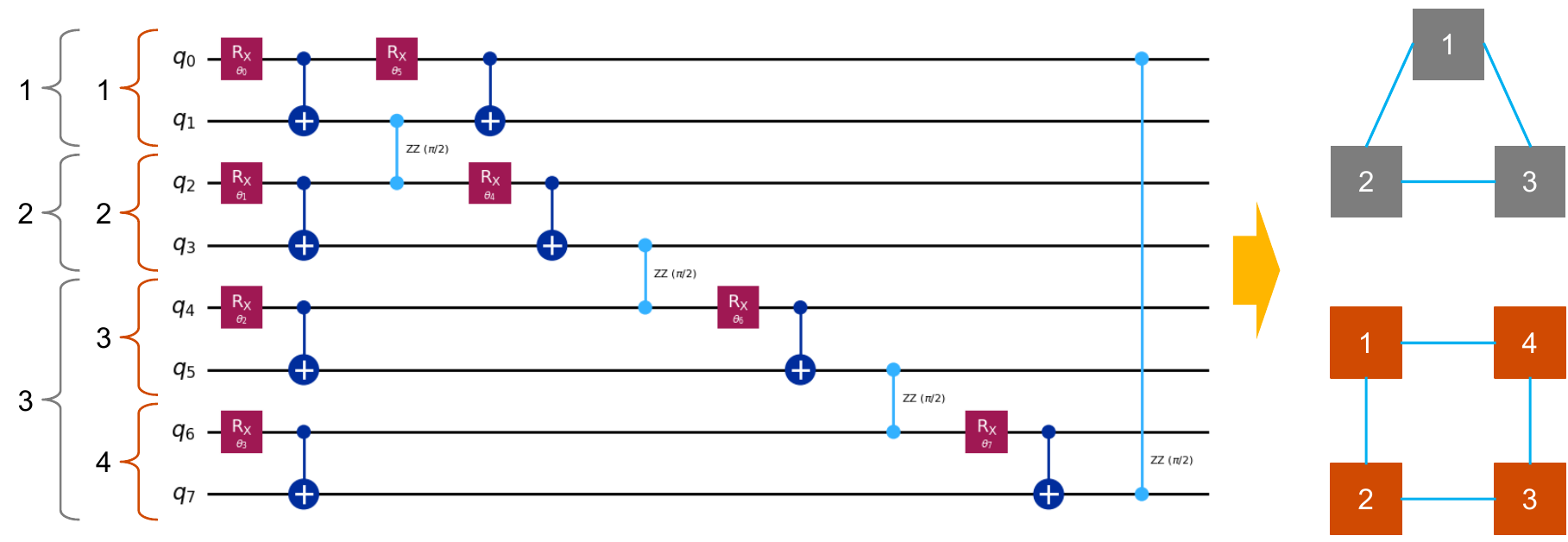}
\end{center}
\caption{An 8-qubit parametric circuit (left) is partitioned to three partitions (upper right) or four partitions (upper left) by cutting three and four $R_{zz}^{}(\theta=\pi/2)$ gates, respectively. The quantum circuit was drawn using Qiskit~\cite{qiskit2024} \texttt{version 1.3.1}.}
\label{fig:circuit}
\end{figure}

\begin{figure}
\begin{center}
\includegraphics[width=8cm]{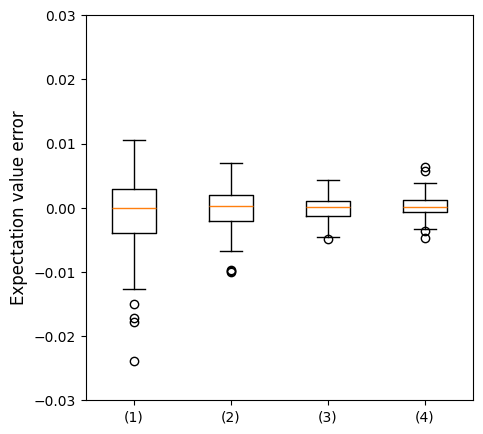}
\end{center}
\caption{Boxplots showing the distribution of expectation value errors for 100 different circuits with randomly chosen parameters in the 8-qubit parametric circuit shown in Figure~\ref{fig:circuit}, evaluated on the observable Pauli $Z^{\otimes 8}_{}$. The errors are defined as the difference between the exact expectation values and those obtained via the circuit cutting algorithm. Labels (1) and (3) correspond to the three-partition case, while (2) and (4) correspond to the four-partition case. Sampling numbers are estimated using eq.~\ref{eq:N_tot} by setting $\epsilon = 0.03$ for (1) and (2), and $\epsilon = 0.01$ for (3) and (4). The observed standard deviations—(1) 0.006, (2) 0.004, (3) 0.002, and (4) 0.002—are all below the respective target errors, confirming the validity of our sampling number estimations.}
\label{fig:error}
\end{figure}

In order to validate our formula, we perform numerical simulations. We consider three and four partitions of an 8-qubit parametric circuit, where the partitions are created by cutting three and four $R_{zz}^{}(\theta=\pi/2)$ gates, respectively (Figure~\ref{fig:circuit}). Note that the $R_{zz}^{}(\theta=\pi/2)$ gate is identical to the CX gate up to local operations.
By randomly choosing parameters in the circuit, we execute 100 different circuits and compute the expectation value of Pauli $Z^{\otimes 8}_{}$. For each circuit, the difference between the exact solution and the solution obtained from the circuit cutting algorithm~\cite{Mitarai_2021} is computed as the expectation value error, and the distributions of these errors are plotted as boxplots (Figure~\ref{fig:error}).
In this figure, (1) and (3) correspond to the results from the three partitions case, and (2) and (4) correspond to those from the four partitions case.
The sampling numbers estimated for $\epsilon = 0.03$ are used in cases (1) and (2), and those for $\epsilon = 0.01$ are used in cases (3) and (4).
The observed standard deviations are (1) $0.006$, (2) $0.004$, (3) $0.002$, and (4) $0.002$, which are smaller than $\epsilon = 0.03$ for (1) and (2) and  $\epsilon = 0.01$ for (3) and (4), as expected.
The sampling numbers estimated using our bound in eq.~\ref{eq:N_tot} and employed in the analysis are (1) $1.2\times 10^6_{}$, (2) $3.2\times 10^6_{}$, (3) $1.1\times 10^7_{}$, and (4) $2.9\times 10^7_{}$.
These numbers may be compared with the estimations obtained by using eq.~\ref{eq:overhead_K}, (1) $8.7 \times 10^6_{}$, (2) $1.0\times 10^8_{}$, (3) $7.8 \times 10^7_{}$, and (4) $9.4 \times 10^8_{}$.
The discrepancy is particularly large in the cases of four partitions, i.e. (2) and (4).
Nevertheless, the numerical results of  (2) and (4) fall within the specified standard deviations, supporting the validity of our sampling number estimations using eq.~\ref{eq:N_tot}. The numerical simulation was performed using Amazon Braket Python SDK~\cite{braket} \texttt{version 1.88.3}.

\section{Partitioning method}\label{sec:method}
In this section, we detail our partitioning method and its implementation. Our method first converts a quantum circuit into an undirected graph with nodes and edges in such a way that each cluster of nodes corresponds to each partition of the circuit.
This is detailed in sec.~\ref{sec:graph}. The subsequent clustering has two steps. The first step is the modularity maximization clustering step. In this step, clusters are merged to increase modularity.
The resulting clusters are further merged in the second step.
These steps are detailed in secs.~\ref{sec:step1} and \ref{sec:step2}, respectively.

\subsection{Doubly weighted graph}\label{sec:graph}

\begin{figure}
\includegraphics[width=16cm]{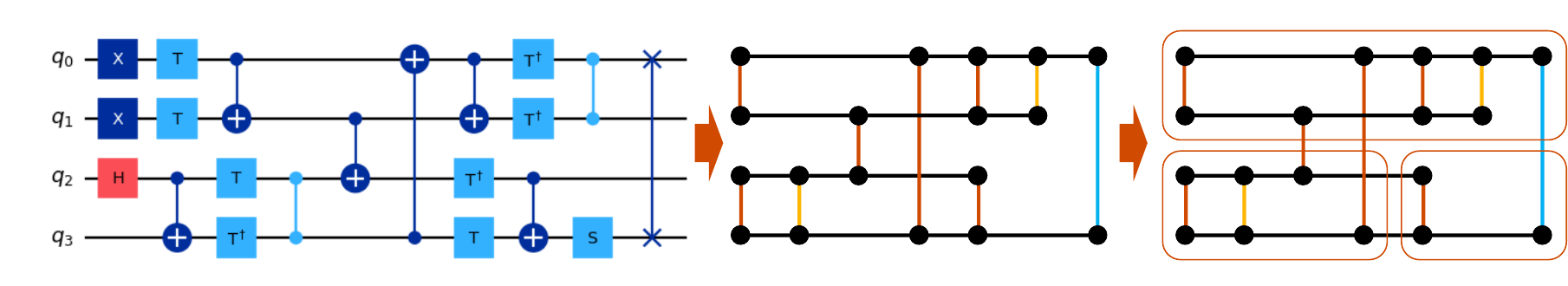}
\caption{A quantum circuit on the left is converted into an undirected graph on the middle which has weights on their edges. Each edge carries the two types of weights at the same time, and the weights of an edge quantifies the sampling overhead caused by cutting the edge.
By performing node clustering through edge cuts, we can achieve circuit partitioning, where one cluster corresponds to one partition. Furthermore, this is realized through combinations of space-like and time-like cuts that can have different weights. An example of three partitions is shown on the right.}
\label{fig:graph}
\end{figure}

We convert a quantum circuit into an weighted graph $G(N, E, w, \hat{w})$ with nodes $N$, edges $E$ and two weight functions $w, \hat{w}: E \to \mathbb{R}$, where each function assigns a weight to each edge.
Assume that we start with an input quantum circuit with 1- or 2-qubit gates. 1-qubit gates are ignored. First, a 2-qubit gate contributes to two nodes at the points where it interacts with the two qubit lines, and one edge between these two nodes. Cutting this edge corresponds to a space-like cut.
Then, adjacent nodes on a common qubit line are connected with an edge, cutting which corresponds to a time-like cut.
By performing node clustering on a graph defined in this way through edge cuts, we can achieve circuit partitioning using a combination of space-like and time-like cuts~\cite{brandhofer2023optimal}, where one cluster corresponds to one partition of the input circuit (see Figure~\ref{fig:graph}).

As shown in eq.~\ref{eq:N_r}, the sampling overhead can be different for each cluster. Based on eq.~\ref{eq:N_r}, we define
\begin{align}
  I_c^{} \coloneq R \Bigl( \prod_{j \in E_c^{}} \kappa_j^{} \Bigr)^2_{}   \prod_{k \in D_c^{}} \tau_k^{},\label{eq:I_c}
\end{align}
where $E_c^{}$ is the set of edges connecting nodes in the cluster $c$ to nodes in other clusters, and $D_c^{}$ is the set of edges connecting nodes in clusters other than the cluster $c$.
Focusing on the largest overhead among the clusters, we define the following quantity as our objective function to minimize and as our metric for evaluating the quality of partitioning solutions,
\begin{subequations}\label{eq:our_metric}
\begin{align}
  L_Q^{} & \coloneq \max_{c \in C}^{} \ln{I_c^{}}  \\
 & =  \max_{c \in C}^{}\Bigl( \ln{R} + \sum_{j \in E_c^{}} \ln{\kappa_j^2} + \sum_{k \in D_c^{}} \ln{\tau_k^{}} \Bigr),
\end{align}
\end{subequations}
where $C$ is the set of clusters.
Since the overhead for each cluster can differ by orders of magnitude, focusing only on the maximum value is a reasonable approach. Moreover, we define
\begin{subequations}\label{eq:LD}
\begin{align}
  r & \coloneq \underset{c \in C} {\operatorname{argmax}} \ln{I_c^{}}, \\
  L_D^{} & \coloneq \sum_{j \in E_r^{}} \ln{\kappa_j^2}.
\end{align}
\end{subequations}
$L_D^{}$ can be a dominant contribution to $L_Q^{}$, therefore it is particularly important to partition the circuit so as to minimize $L_D^{}$.

We define the weight functions $w$ and $\hat{w}$ for edge $i \in E$ as
\begin{subequations}\label{eq:weight}
\begin{align}
  w_i^{} & \coloneq \ln{\kappa_i^2}, \label{eq:weight_1}\\
  \hat{w}_i^{} & \coloneq \ln{\tau_i^{}}.  \label{eq:weight_2}
\end{align}
\end{subequations}
Namely, each edge carries the two types of weights at the same time. By defining weights in this way, computing $L_Q^{}$ and $L_D^{}$ is as straightforward as counting the number of cut edges or computing the cluster degrees, as done in refs.~\cite{tang2021cutqc,tang2022scaleqc}.
Throughout the following discussion, the term "weight" primarily refers to the weight $w$ defined by eq.~\ref{eq:weight_1}.

\subsection{Qubit-constrained modularity clustering step}\label{sec:step1}

The first step of node clustering on our weighted graphs is modularity maximization~\cite{Clauset_2004}, where clusters are subsequently merged to increase modularity defined as~\cite{newman2004analysis, Clauset_2004}
\begin{subequations}\label{eq:modularity}
\begin{align}
  q_c^{} & \coloneq \frac{M_c^{}}{m} - \biggl(\frac{\Sigma_c^{}}{2m} \biggr)^2_{}, \label{eq:modularity_1}\\
  Q & \coloneq \sum_{c \in C} q_c^{},
\end{align}
\end{subequations}
where $m$ is the sum of the weights of all the edges, $m \coloneq \sum_i^{\mathrm{all}} w_i^{}$, $M_c^{}$ is the sum of the weights of edges connecting nodes within the cluster $c$ (each edge is counted only once), and $\Sigma_c^{}$ is the sum of the weights of the edges attached to nodes within the cluster $c$ (double counting is possible if both endpoints of an edge belong to the same cluster).
It is straightforward to confirm that
\begin{subequations}\label{eq:sigma_c}
\begin{align}
  \Sigma_c^{} & = 2 M_c^{} + \sum_{i \in E_c^{}} w_i^{} \\
  & = 2 M_c^{} + \sum_{i \in E_c^{}} \ln{\kappa_j^2} , \label{eq:k_c}
\end{align}
\end{subequations}
where eq.~\ref{eq:weight_1} is used at the last equal.
Modularity maximization clusters nodes in a graph in such a way that increases the first term and decreases the second term in eq.~\ref{eq:modularity_1}.
Since eq.~\ref{eq:k_c} shows that the second term in eq.~\ref{eq:modularity_1} includes $\ln{\kappa_j^2}$, modularity maximization with our weight definition in eq.~\ref{eq:weight} can reduce $L_D^{}$, which can be the dominant contribution to our objective function $L_Q^{}$ (see eqs.~\ref{eq:LD} ~\ref{eq:our_metric}).
This observation is one of our motivations to adopt modularity maximization.
Another motivation is that there exists a simple heuristic algorithm that is known to quickly produce high-quality solutions~\cite{blondel2008fast}.
Our algorithm of modularity maximization is based on ref.~\cite{blondel2008fast} and its implementation is influenced by refs.~\cite{hagberg2008exploring, networkx_github}.

Our weighted graph $G(N, E, w, \hat{w})$ and the maximum number of qubits in devices, $D$, are inputs to our algorithm.
First, we assign each node $i \in N$ to its own cluster $c(i)$, resulting in $|N|$ clusters.
Then, we pick a node $i \in N$.
Next, we select a cluster $c^{\prime}_{} \in C(i)$ where $C(i)$ is a set of clusters containing nodes connected to the node $i$ via edges.
Assuming that the node $i$ is moved from $c(i)$ to $c^{\prime}_{}$, we calculate the resulting number of qubits in $c^{\prime}_{}$.
If that number is less than or equal to $D$, we  calculate the gain of modularity by moving $i$ from $c(i)$ to $c^{\prime}_{}$.
This calculation is performed for all the other $c^{\prime}_{} \in C(i)$.
The node $i$ is actually moved to $c^{\prime}_{}$ that yields the maximum gain.
If the maximum gain is zero or negative, the node $i$ remains in $c(i)$.
This operation is repeated for all the other $i \in N$.
The above is repeated as long as node movements occur.
This constitutes the first phase.

In the second phase, a new weighted graph is generated by contracting the clusters obtained in the first phase (each containing at least one node) into supernodes.
The sum of the weights of all edges connecting two clusters becomes the weight of the new edge connecting the corresponding two supernodes.
The sum of the weights of edges connecting nodes within a cluster is retained as the weight of the self-loop edge of the corresponding supernode.

The new weighted graph will be used as the input to the upcoming first phase.
These two phases together are referred to as a "pass". This pass is repeated as long as node movements (i.e. increases in modularity) occur during the first phase.
Note that there is arbitrariness in the order of selecting node $i \in N$ in the first phase.
In our numerical studies, we take the descending order of the sum of the weights of the edges connected to node $i \in N$.
However, since our method is very fast as shown in Sec.~\ref{sec:qasmbench}, it is an option to run the random-order approach multiple times and select the best outcome in terms of $L_Q^{}$.

\begin{figure}
\begin{center}
\includegraphics[width=10cm]{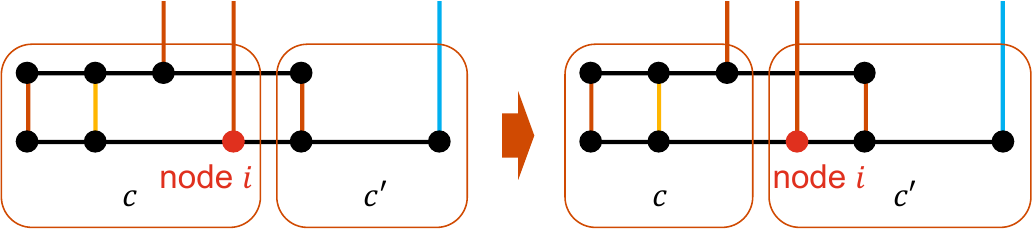}
\end{center}
\caption{An example of moving a node $i$ from its cluster $c$ to an adjacent cluster $c^{\prime}_{}$. This operation is part of the modularity maximization step, where the gain of modularity can be efficiently computed using only the local information from the clusters $c$ and $c^{\prime}_{}$.}
\label{fig:delta_q}
\end{figure}

\begin{algorithm}
\caption{Qubit-Constrained Modularity Clustering (Step 1)}
\label{alg:alg1}
\KwIn{Graph $G(N, E, w, \hat{w})$, maximum number of qubits per cluster $D$}
\KwOut{Contracted graph $G^{\prime}_{}$}
\ForEach{$i \in N$}{
  Assign $i$ to its own cluster $c(i)$\;
}
repeat $\gets$ \textbf{true}\;
\While{repeat is \textbf{true}}{
  repeat $\gets$ \textbf{false}\;
  \tcp{Phase 1: Local modularity maximization}
  \ForEach{node $i \in N$ (in weighted order)}{
    $current\_cluster \gets c(i)$\;
    $best\_gain \gets 0$\;
    $best\_cluster \gets current\_cluster$\;
    Compute $\Delta Q_{\mathrm{remove}}(i)$\;
    $C(i) \gets$ set of clusters connected to $i$ via edges\;
    \ForEach{$c^{\prime}_{} \in C(i)$}{
      \If{moving $i$ to $c^{\prime}_{}$ does not exceed qubit limit $D$}{
        Compute $\Delta Q_{\mathrm{add}}(i, c^{\prime}_{})$\;
        $gain \gets \Delta Q_{\mathrm{remove}}(i) + \Delta Q_{\mathrm{add}}(i, c^{\prime}_{})$\;

        \If{$gain > best\_gain$}{
          $best\_gain \gets gain$\;
          $best\_cluster \gets c^{\prime}_{}$\;
        }
      }
    }
    \If{$best\_cluster \neq current\_cluster$}{
      $c(i) \gets best\_cluster$\;
      repeat $\gets$ \textbf{true}\;
    }
  }
  \tcp{Phase 2: Graph contraction}
  \If{repeat is \textbf{true}}{
    Contract each cluster into a supernode\;
    \tcp{Edge weight between supernodes = sum of weights of inter-cluster edges}
    \tcp{Self-loop weight = sum of weights of intra-cluster edges}
    Generate a new graph $G^{\prime}_{}$ from supernodes (i.e. $N, E, w, \hat{w}$ are updated)\;
    Reinitialize $c(i)$ for the new graph\;
  }
}
\Return{$G^{\prime}_{}$}\;
\end{algorithm}

The gain of modularity can be efficiently calculated  from~\cite{blondel2008fast}
\begin{align}
  \Delta Q(i, c^{\prime}_{}) = \Delta Q_{\mathrm{remove}}^{}(i) + \Delta Q_{\mathrm{add}}^{}(i, c^{\prime}_{}), \label{eq:mod1}
\end{align}
where $\Delta Q_{\mathrm{remove}}^{}(i)$ is the gain of modularity by removing the node $i$ from the cluster $c(i)$:
\begin{align}
  \Delta Q_{\mathrm{remove}}^{}(i) = - \frac{k_{i, c(i)}^{}}{m} + \frac{k_i^{}(\Sigma_{c(i)}^{}-k_i^{})}{2m^2_{}},\label{eq:mod2}
\end{align}
and $\Delta Q_{\mathrm{add}}^{}(i, c^{\prime}_{})$ is the gain of modularity by adding the node $i$ to the cluster $c^{\prime}_{}$:
\begin{align}
  \Delta Q_{\mathrm{add}}^{}(i, c^{\prime}_{}) = \frac{k_{i, c^{\prime}_{}}^{}}{m} - \frac{k_i^{} \cdot \Sigma_{c^{\prime}_{}}^{}}{2m^2_{}}, \label{eq:mod3}
\end{align}
Here $k_{i, c}^{}$ is the sum of the weights of edges connecting the node $i$ to nodes within the cluster $c$, and $k_i^{}$ is the sum of the weights of the edges attached to the node $i$ (including the self-loop edges). In Figure~\ref{fig:delta_q}, we show an example of moving a node $i$ from its cluster $c(\coloneq c(i))$ to an adjacent cluster $c^{\prime}_{}$. For this simple example, we can easily find (refer to Figure~\ref{fig:graph} for the correspondence between the graph and the quantum circuit, and to Table~\ref{tbl:kappa_tau} for the weights of edges)
\begin{align}
k_{i, c}^{} = \ln{16}, \ \ k_{i, c^{\prime}_{}}^{} = \ln{16}, \ \ k_i^{} = 2\cdot \ln{16} + \ln{9}, \nonumber \\
\Sigma_c^{} = 10 \cdot \ln{16} + 6 \cdot \ln{9}, \ \ \Sigma_{c^{\prime}_{}}^{} = 4 \cdot \ln{16} + 2 \cdot \ln{9} + \ln{49}. \ \
\end{align}
By inputting these values into eqs.~\ref{eq:mod1} \ref{eq:mod2} \ref{eq:mod3}, the gain of modularity can be calculated using only the information from the cluster $c$ and $c^{\prime}$ as
\begin{align}
  \Delta Q(i, c^{\prime}_{}) = \frac{(2\cdot \ln{16} + \ln{9})(4\cdot \ln{16} + 3 \cdot \ln{9} - \ln{49})}{2 m^2_{}}
\end{align}
where $m$ denotes the sum of the weights of all the edges and is a constant during the algorithm. We show the flow of the algorithm in Algorithm~\ref{alg:alg1}.

The use of modularity maximization for identifying cutting locations has already been validated in ref.~\cite{kan2024scalable}.
However, ref.~\cite{kan2024scalable} considers only time-like cutting. In our method, by considering space-like cutting as well as time-like cutting, the size of a graph becomes twice as large as that in ref.~\cite{kan2024scalable}  in terms of the number of nodes. The scalability of modularity maximization with respect to graph size demonstrates its strength when dealing with such large-scale graphs.

\subsection{$L_Q^{}$ minimization clustering step}\label{sec:step2}
The modularity maximization algorithm~\cite{blondel2008fast} adopted in the first step is primarily intended to perform community detection, rather than to directly minimize our objective function $L_Q^{}$ in eq.~\ref{eq:our_metric} (although it is expected to contribute to reduction of it via reduction of $L_D^{}$ as shown below eq.~\ref{eq:k_c}). Therefore, the second step takes the result of the first step as input and further merges clusters. The merging is performed so as to reduce the value of $L_Q^{}$.

\begin{algorithm}
\caption{$L_Q^{}$ Minimization Clustering (Step 2)}
\label{alg:alg2}
\KwIn{Contracted graph $G(N, E, w, \hat{w})$ (output of Algorithm 1), maximum number of qubits per cluster $D$}
\KwOut{Further contracted graph $G^{\prime}_{}$}
\ForEach{$i \in N$}{
  Assign $i$ to its own cluster $c(i)$\;
}
repeat $\gets$ \textbf{true}\;
$G^{\prime}_{} \gets G$\;
\While{repeat is \textbf{true}}{
  repeat $\gets$ \textbf{false}\;
  Compute $L_Q^{}$ for $G^{\prime}_{}$\;
  \tcp{Phase 1: $L_Q^{}$ Minimization}
  \ForEach{node $i \in N$ (in random or weighted order)}{
    $current\_cluster \gets c(i)$\;
    $best\_cluster \gets current\_cluster$\;
    Compute $\ln{I_c^{}}$ for $current\_cluster$ after removing $i$\;
    \If{$\ln{I_c^{}} \leq L_Q^{}$}{
      $C(i) \gets$ set of clusters connected to $i$ via edges\;
      \ForEach{$c^{\prime}_{} \in C(i)$}{
        \If{moving $i$ to $c'$ does not exceed qubit limit $D$}{
          Compute $\ln{I_{c^{\prime}_{}}^{}}$ and $dw_{i, c^{\prime}_{}}^{}$ after adding $i$ to $c^{\prime}_{}$\;
          \If{$\ln{I_{c^{\prime}_{}}^{}} < L_Q^{}$}{
            $L_Q^{} \gets l_{c^{\prime}_{}}^{}$\;
            $best\_cluster \gets c^{\prime}_{}$\;
          }
          \ElseIf{$\ln{I_{c^{\prime}_{}}^{}} = L_Q^{}$ \textbf{and} $dw_{i, c^{\prime}_{}}^{} < 0$}{
            $best\_cluster \gets c^{\prime}_{}$\;
          }
        }
      }
      \If{$best\_cluster \neq current\_cluster$}{
        $c(i) \gets best\_cluster$\;
        repeat $\gets$ \textbf{true}\;
      }
    }
  }
  \tcp{Phase 2: Graph Contraction}
  \If{repeat is \textbf{true}}{
    Contract each cluster into a supernode\;
    \tcp{Edge weight between supernodes = sum of weights of inter-cluster edges}
    \tcp{Self-loop weight = sum of weights of intra-cluster edges}
    Generate a new graph $G^{\prime}_{}$ from supernodes (i.e. $N, E, w, \hat{w}$ are updated)\;
    Reinitialize $c(i)$ for the new graph\;
  }
}
\Return{$G^{\prime}_{}$}\;
\end{algorithm}

The algorithm proceeds in almost the same manner as the first step. It consists of two phases, with the second phase being exactly the same. A "pass", which is the combination of two phases, is repeated in the same way. Differences lie in the first phase. At the beginning of each pass, the metric $L_Q^{}$ is computed. First, a node $i \in N$ is picked up in the same way as the first step. Assuming that the node $i$ is removed from $c(i)$, the resulting $\ln{I_c^{}}$ in eq.~\ref{eq:I_c} is calculated for $c(i)$. If this value is less than or equal to $L_Q^{}$, we proceed to  select a cluster $c^{\prime}_{} \in C(i)$ and compute the resulting $\ln{I_{c^{\prime}_{}}^{}}$ for $c^{\prime}_{}$ assuming that the node $i$ is added to $c^{\prime}_{}$.
 Furthermore, we compute the change in the sum of the weights of all inter-cluster edges resulting from moving the node $i$ from $c(i)$ to $c^{\prime}_{}$ (denoted by $dw_{i, c^{\prime}_{}}^{}$).
If $\ln{I_{c^{\prime}_{}}^{}} < L_Q^{}$, or if $\ln{I_{c^{\prime}_{}}^{}} = L_Q^{}$ {\it and} $dw_{i, c^{\prime}_{}}^{} < 0$, then the node $i$ is moved to $c^{\prime}_{}$, and $L_Q^{}$ is updated to $\ln{I_{c^{\prime}_{}}^{}}$. In all the other cases, the node $i$ remains  in $c(i)$. Naturally, it is also required that the number of qubits resulting from adding the node $i$ to $c^{\prime}_{}$ does not exceed $D$. This process is applied repeatedly and subsequently for all other clusters $c^{\prime}_{} \in C(i)$ and nodes $i \in N$, as long as node movements occur.
We show the flow of the algorithm in Algorithm~\ref{alg:alg2}.

One of the advantages of our method, which consists of step 1 and step 2, is that the optimal number of partitions is also automatically determined during the optimization process.
In contrast, in many of graph partitioning algorithms such as mixed-integer programming (MIP) models, the number of partitions must be specified as input. For example, in ref.~\cite{tang2022scaleqc}, which adopts a MIP model, their solver is run for all possible partition numbers from 2 to $n$ for a $n$-qubit circuit in order to determine the optimal number of partitions. This approach may become computationally expensive for circuits with a large number of qubits.

\section{Application to the QASMBench}\label{sec:qasmbench}

\begin{table}
\begin{tabular}{p{7em}p{3em}p{3em}p{3em}p{3em}p{3em}p{3em}p{3em}p{3em}}
\toprule
circuit & \multicolumn{2}{l}{$L_Q^{}$} & \multicolumn{2}{l}{$L_D^{}$} & \multicolumn{2}{l}{$R$} & \multicolumn{2}{l}{Time[sec]} \\
\cmidrule(r){2-3}
\cmidrule(r){4-5}
\cmidrule(r){6-7}
\cmidrule(r){8-9}
 & Step 1 & Step 2 & Step 1 & Step 2 & Step 1 & Step 2 & Step 1 & Step 2 \\
\midrule
ising\sus n34\sus D30 & 17.33 & 3.47 & 5.55 & 2.77 & 16 & 2 & 0.0025 & 0.0005 \\
ising\sus n66\sus D30 & 18.08 & 6.64 & 5.55 & 5.55 & 17 & 3 & 0.0051 & 0.0007 \\
ising\sus n66\sus D50 & 18.08 & 3.47 & 5.55 & 2.77 & 17 & 2 & 0.0056 & 0.0009 \\
ising\sus n98\sus D30 & 23.28 & 7.62 & 5.55 & 5.55 & 24 & 4 & 0.0080 & 0.0014 \\
ising\sus n98\sus D50 & 23.28 & 6.64 & 5.55 & 5.55 & 24 & 3 & 0.0080 & 0.0010 \\
ising\sus n98\sus D70 & 23.28 & 3.47 & 5.55 & 2.77 & 24 & 2 & 0.0080 & 0.0008 \\
ising\sus n420\sus D30 & 43.46 & 18.83 & 5.55 & 5.55 & 52 & 18 & 0.0887 & 0.0131 \\
ising\sus n420\sus D50 & 43.46 & 11.90 & 5.55 & 5.55 & 52 & 9 & 0.0342 & 0.0072 \\
ising\sus n420\sus D70 & 43.46 & 10.26 & 5.55 & 5.55 & 52 & 7 & 0.0340 & 0.0057 \\
\bottomrule
\end{tabular}
\caption{For the Ising circuits from QASMBench, the outputs of the first step in our algorithm are compared with those of the second step which takes the output of the first step as its input.
Each circuit is labeled using the format: ising\sus n(the number of qubits in the circuit)\sus D(the maximum number of qubits allowed per partition). We measure our metric $L_Q^{}$ in eq.~\ref{eq:our_metric}, $L_D^{}$ in eq.~\ref{eq:LD},  the number of partitions (i.e. clusters) $R$, and the execution time.}
\label{tbl:step1step2}
\end{table}

In order to verify the validity of our method, we apply it on a number of benchmark quantum circuits and compare the results with those obtained using \texttt{find\_cuts} function from Qiskit-addon-cutting (QAC)~\cite{qiskit-addon-cutting} \texttt{version 0.10.0}, which is a representative tool that integrates the functions required for circuit cutting.
As benchmark quantum circuits, we adopt the circuits from the large-scale category of QASMBench~\cite{li2023qasmbench}. Only circuits in which all entangling gates have been transpiled into CX gates, as given by \texttt{xxx\_transpiled.qasm}, are adopted.
However, circuits containing mid-circuit measurements (e.g., \texttt{cc\_nyyy\_transpiled.qasm}) and those that are excessively large (e.g., \texttt{bwt\_nyyy\_transpiled.qasm}) are excluded.

To first verify that our algorithm behaves as intended, we compare the outputs of the first step with those of the second step which takes the output of the first step as its input. Using the Ising circuits from QASMBench as an example, we measure the metric $L_Q^{}$ in eq.~\ref{eq:our_metric}, $L_D^{}$ in eq.~\ref{eq:LD}, the number of partitions $R$, and the execution time (see Table~\ref{tbl:step1step2}).
We can observe that the value of $L_D^{}$ is nearly minimized in the first step, and in some cases, no further reduction is observed in the second step. This may be expected from the property of modularity maximization as explained below eq.~\ref{eq:sigma_c}.
In the second step, $L_Q^{}$ is meaningfully reduced while  $L_D^{}$ is preserved or reduced. This indicates that the second step reduces $R$ and minimizes the contributions other than $L_D^{}$.
It is also found that, when $R$ is large,  $L_D^{}$ is not necessarily the dominant contribution to $L_Q^{}$.
The execution time is generally quite fast. In particular, the second step tends to be faster than the first step, as it starts with a smaller number of initial clusters.

For a larger set of circuits (83 in total) from QASMBench, we compare the results of this work's algorithm (TWA) with those of QAC.
Let us define $n_{\mathrm{space}}^{}$ ($n_{\mathrm{time}}^{}$) as the number of space-like (time-like) edges in $E_r^{}$. Since all entangling gates in the circuits are already transpiled into CX gates, $L_D^{}$ in eq.~\ref{eq:LD} can be expressed as
\begin{align}
  L_D^{}  =  n_{\mathrm{space}}^{} \cdot \ln{9} + n_{\mathrm{time}}^{} \cdot \ln{16}. \label{eq:LD_2}
\end{align}
 We also introduce $n_{\mathrm{tot, space}}^{}$, the total number of space-like cuts, and $n_{\mathrm{tot, time}}^{}$, the total number of space-like cuts, and define
\begin{align}
  L_{\mathrm{tot}}^{} \coloneq n_{\mathrm{tot, space}}^{} \cdot \ln{9} + n_{\mathrm{tot, time}}^{} \cdot \ln{16}.
\end{align}
Note that QAC partitions the circuit with the objective of minimizing the quantity corresponding to  $L_{\mathrm{tot}}^{}$.

The two approaches are compared in terms of $L_Q^{}$, $n_{\mathrm{space}}^{}$, $n_{\mathrm{time}}^{}$, $L_{\mathrm{tot}}^{}$, $R$ and the execution time (see Table~\ref{tbl:all}).
Regarding $L_Q^{}$, we can observe that the results of TWA are equal to or smaller (i.e. better) than those of QAC in all but 6 out of the 83 circuits.
As for $L_{\mathrm{tot}}^{}$, TWA performs worse than QAC only in 6 circuits.
Furthermore, QAC encountered overflow errors in a number of the circuits, which are denoted by \texttt{-|-}.
It is noteworthy that, although the quantity corresponding to  $L_{\mathrm{tot}}^{}$ is the objective function of QAC, TWA still achieves the better results with respect to $L_{\mathrm{tot}}^{}$ on average.
With respect to execution time, TWA significantly outperforms QAC by achieving a speedup ranging from $\sim 2 \times$ to $\sim 3080 \times$ depending on the circuit with an average speedup of $\sim 610 \times$.

{\footnotesize
\begin{longtable}{lllllllllllllllllll}
\caption{For a set of circuits from the large-scale category of QASMBench, we compare the results of this work's algorithm (TWA) with those of Qiskit-addon-cutting (QAC) with respect to a number of metrics. Each circuit is labeled using the format: (circuit name)\sus n(the number of qubits in the circuit)\sus D(the maximum number of qubits allowed per partition). The symbol  \texttt{-|-} indicates that QAC encountered an overflow error. $\prescript{a}{}{}$Our simulation found an error in QAC that the number of partitions is unnecessarily increased.}
\label{tbl:all}
\\
\toprule
circuit & \multicolumn{2}{l}{$L_Q^{}$} & \multicolumn{2}{l}{$n_{\mathrm{space}}^{}/n_{\mathrm{time}}^{}$} & \multicolumn{2}{l}{$L_{\mathrm{tot}}^{}$}  & \multicolumn{2}{l}{$R$} & \multicolumn{2}{l}{Time[sec]} \\
\cmidrule(r){2-3}
\cmidrule(r){4-5}
\cmidrule(r){6-7}
\cmidrule(r){8-9}
\cmidrule(r){10-11}
 & TWA & QAC & TWA & QAC & TWA & QAC & TWA & QAC & TWA & QAC \\
\midrule
\endfirsthead
\toprule
circuit & \multicolumn{2}{l}{$L_Q^{}$} & \multicolumn{2}{l}{$n_{\mathrm{space}}^{}/n_{\mathrm{time}}^{}$} & \multicolumn{2}{l}{$L_{\mathrm{tot}}^{}$}  & \multicolumn{2}{l}{$R$} & \multicolumn{2}{l}{Time[sec]} \\
\cmidrule(r){2-3}
\cmidrule(r){4-5}
\cmidrule(r){6-7}
\cmidrule(r){8-9}
\cmidrule(r){10-11}
 & TWA & QAC & TWA & QAC & TWA & QAC & TWA & QAC & TWA & QAC \\
\midrule
\endhead
\midrule
\multicolumn{11}{r}{Continued on next page} \\
\midrule
\endfoot
\bottomrule
\endlastfoot
adder\sus n64\sus D30 & 5.5 & 5.5 & 2/0 & 2/0 & 4.4 & 4.4 & 3 & 3 & 0.04 & 0.80 \\
adder\sus n64\sus D50 & 2.9 & 2.9 & 1/0 & 1/0 & 2.2 & 2.2 & 2 & 2 & 0.03 & 0.05 \\
adder\sus n118\sus D30 & 6.8 & 54 & 2/0 & 22/0 & 8.8 & 72 & 5 & 4 & 0.06 & 50.9 \\
adder\sus n118\sus D50 & 5.5 & 5.5 & 2/0 & 2/0 & 4.4 & 4.4 & 3 & 3 & 0.06 & 2.15 \\
adder\sus n118\sus D70 & 2.9 & 2.9 & 1/0 & 1/0 & 2.2 & 2.2 & 2 & 2 & 0.06 & 0.10 \\
adder\sus n433\sus D30 & 12 & 104 & 2/0 & 22/0 & 32 & 338 & 16 & 15 & 0.24 & 74.6 \\
adder\sus n433\sus D50 & 9.5 & 77 & 2/0 & 22/0 & 19 & 193 & 10 & 9 & 0.29 & 87.7 \\
adder\sus n433\sus D70 & 8.0 & 68 & 2/0 & 22/0 & 13 & 145 & 7 & 7 & 0.23 & 85.6 \\
bv\sus n70\sus D30 & 3.5 & 6.3 & 0/1 & 0/1 & 2.8 & 2.8 & 2 & $35^a_{}$ & $<0.01$ & 0.04 \\
bv\sus n140\sus D30 & 6.6 & 9.8 & 0/2 & 0/2 & 5.5 & 5.5 & 3 & $70^a_{}$ & $<0.01$ & 2.54 \\
bv\sus n140\sus D50 & 3.5 & 7.0 & 0/1 & 0/1 & 2.8 & 2.8 & 2 & $69^a_{}$ & $<0.01$ & 0.09 \\
bv\sus n140\sus D70 & 3.5 & 7.0 & 0/1 & 0/1 & 2.8 & 2.8 & 2 & $69^a_{}$ & $<0.01$ & 0.16 \\
bv\sus n280\sus D30 & 10 & 275 & 0/2 & 123/0 & 16 & 270 & 7 & 251 & 0.01 & 9.54 \\
bv\sus n280\sus D50 & 7.6 & 231 & 0/2 & 103/0 & 8.3 & 226 & 4 & 231 & 0.01 & 22.0 \\
bv\sus n280\sus D70 & 6.6 & 187 & 0/2 & 83/0 & 5.5 & 182 & 3 & 211 & 0.01 & 22.2 \\
cat\sus n65\sus D30 & 5.5 & 5.5 & 2/0 & 2/0 & 4.4 & 4.4 & 3 & 3 & $<0.01$ & 0.11 \\
cat\sus n65\sus D50 & 2.9 & 2.9 & 1/0 & 1/0 & 2.2 & 2.2 & 2 & 2 & $<0.01$ & 0.01 \\
cat\sus n130\sus D30 & 6.8 & 6.8 & 2/0 & 2/0 & 8.8 & 8.8 & 5 & 5 & 0.01 & 16.4 \\
cat\sus n130\sus D50 & 5.5 & 5.5 & 2/0 & 2/0 & 4.4 & 4.4 & 3 & 3 & 0.01 & 0.31 \\
cat\sus n130\sus D70 & 2.9 & 2.9 & 1/0 & 1/0 & 2.2 & 2.2 & 2 & 2 & 0.01 & 0.01 \\
cat\sus n260\sus D30 & 10 & 9.0 & 2/0 & 2/0 & 21 & 17 & 11 & 9 & 0.02 & 20.6 \\
cat\sus n260\sus D50 & 7.4 & 7.4 & 2/0 & 2/0 & 10 & 10 & 6 & 6 & 0.01 & 33.9 \\
cat\sus n260\sus D70 & 6.2 & 6.2 & 2/0 & 2/0 & 6.6 & 6.6 & 4 & 4 & 0.01 & 32.4 \\
ghz\sus n78\sus D30 & 5.5 & 5.5 & 2/0 & 2/0 & 4.4 & 4.4 & 3 & 3 & $<0.01$ & 0.11 \\
ghz\sus n78\sus D50 & 2.9 & 2.9 & 1/0 & 1/0 & 2.2 & 2.2 & 2 & 2 & $<0.01$ & 0.01 \\
ghz\sus n78\sus D70 & 2.9 & 2.9 & 1/0 & 1/0 & 2.2 & 2.2 & 2 & 2 & $<0.01$ & 0.01 \\
ghz\sus n127\sus D30 & 6.8 & 6.8 & 2/0 & 2/0 & 8.8 & 8.8 & 5 & 5 & 0.01 & 16.6 \\
ghz\sus n127\sus D50 & 5.5 & 5.5 & 2/0 & 2/0 & 4.4 & 4.4 & 3 & 3 & 0.01 & 0.30 \\
ghz\sus n127\sus D70 & 2.9 & 2.9 & 1/0 & 1/0 & 2.2 & 2.2 & 2 & 2 & 0.01 & 0.01 \\
ghz\sus n255\sus D30 & 10 & 9.0 & 2/0 & 2/0 & 21 & 17 & 11 & 9 & 0.01 & 19.9 \\
ghz\sus n255\sus D50 & 7.4 & 7.4 & 2/0 & 2/0 & 10 & 10 & 6 & 6 & 0.01 & 33.6 \\
ghz\sus n255\sus D70 & 6.2 & 6.2 & 2/0 & 2/0 & 6.6 & 6.6 & 4 & 4 & 0.01 & 32.2 \\
ising\sus n66\sus D30 & 6.6 & 6.6 & 0/2 & 0/2 & 5.5 & 5.5 & 3 & 3 & 0.01 & 0.39 \\
ising\sus n66\sus D50 & 3.5 & 3.5 & 0/1 & 0/1 & 2.8 & 2.8 & 2 & 2 & 0.01 & 0.02 \\
ising\sus n98\sus D30 & 7.6 & 10 & 0/2 & 4/0 & 8.3 & 13 & 4 & 4 & 0.01 & 16.1 \\
ising\sus n98\sus D50 & 6.6 & 3.5 & 0/2 & 0/1 & 5.5 & 2.8 & 3 & 2 & 0.01 & 0.03 \\
ising\sus n98\sus D70 & 3.5 & 3.5 & 0/1 & 0/1 & 2.8 & 2.8 & 2 & 2 & 0.01 & 0.03 \\
ising\sus n420\sus D30 & 18 & 20 & 0/2 & 4/0 & 47 & 57 & 18 & 14 & 0.05 & 17.9 \\
ising\sus n420\sus D50 & 11 & 15 & 0/2 & 4/0 & 22 & 35 & 9 & 9 & 0.05 & 27.3 \\
ising\sus n420\sus D70 & 10 & 13 & 0/2 & 4/0 & 16 & 21 & 7 & 6 & 0.05 & 13.7 \\
knn\sus n67\sus D30 & 6.6 & 6.6 & 0/2 & 0/2 & 5.5 & 5.5 & 3 & 3 & 0.01 & 0.45 \\
knn\sus n67\sus D50 & 3.5 & 3.5 & 0/1 & 0/1 & 2.8 & 2.8 & 2 & 2 & 0.01 & 0.04 \\
knn\sus n129\sus D30 & 8.5 & 443 & 0/2 & 200/0 & 11 & 439 & 5 & 51 & 0.03 & 6.58 \\
knn\sus n129\sus D50 & 6.6 & 6.6 & 0/2 & 0/2 & 5.5 & 5.5 & 3 & 3 & 0.03 & 1.37 \\
knn\sus n129\sus D70 & 3.5 & 3.5 & 0/1 & 0/1 & 2.8 & 2.8 & 2 & 2 & 0.03 & 0.08 \\
knn\sus n341\sus D30 & 15 & 1376 & 0/2 & 624/0 & 36 & 1371 & 14 & 157 & 0.08 & 7.53 \\
knn\sus n341\sus D50 & 11 & 1288 & 0/2 & 584/0 & 19 & 1283 & 8 & 147 & 0.12 & 26.6 \\
knn\sus n341\sus D70 & 9.4 & 1200 & 0/2 & 544/0 & 13 & 1195 & 6 & 137 & 0.07 & 30.6 \\
qft\sus n63\sus D30 & 318 & \texttt{-|-} & 0/60 & \texttt{-|-} & 765 & \texttt{-|-} & 13 & \texttt{-|-} & 0.26 & \texttt{-|-} \\
qft\sus n63\sus D50 & 246 & \texttt{-|-} & 0/81 & \texttt{-|-} & 307 & \texttt{-|-} & 4 & \texttt{-|-} & 0.27 & \texttt{-|-} \\
qft\sus n160\sus D30 & 1020 & \texttt{-|-} & 0/70 & \texttt{-|-} & 3482 & \texttt{-|-} & 52 & \texttt{-|-} & 1.02 & \texttt{-|-} \\
qft\sus n160\sus D50 & 643 & \texttt{-|-} & 0/102 & \texttt{-|-} & 1713 & \texttt{-|-} & 17 & \texttt{-|-} & 2.94 & \texttt{-|-} \\
qft\sus n160\sus D70 & 481 & \texttt{-|-} & 0/96 & \texttt{-|-} & 1120 & \texttt{-|-} & 10 & \texttt{-|-} & 1.33 & \texttt{-|-} \\
qft\sus n320\sus D30 & 2120 & \texttt{-|-} & 0/70 & \texttt{-|-} & 7879 & \texttt{-|-} & 115 & \texttt{-|-} & 2.80 & \texttt{-|-} \\
qft\sus n320\sus D50 & 1169 & \texttt{-|-} & 0/116 & \texttt{-|-} & 3698 & \texttt{-|-} & 36 & \texttt{-|-} & 3.14 & \texttt{-|-} \\
qft\sus n320\sus D70 & 704 & \texttt{-|-} & 0/102 & \texttt{-|-} & 1960 & \texttt{-|-} & 16 & \texttt{-|-} & 3.10 & \texttt{-|-} \\
qugan\sus n71\sus D30 & 32 & 448 & 0/6 & 202/0 & 72 & 461 & 9 & 3 & 0.03 & 13.7 \\
qugan\sus n71\sus D50 & 9.0 & 286 & 0/3 & 130/0 & 8.3 & 285 & 2 & 2 & 0.22 & 30.8 \\
qugan\sus n71\sus D70 & 9.0 & 22 & 0/3 & 10/0 & 8.3 & 21 & 2 & 2 & 0.04 & 26.5 \\
qugan\sus n111\sus D30 & 22 & 659 & 0/6 & 282/0 & 33 & 830 & 5 & 4 & 0.05 & 14.7 \\
qugan\sus n111\sus D50 & 17 & 711 & 0/6 & 322/0 & 16 & 725 & 3 & 3 & 0.05 & 50.7 \\
qugan\sus n111\sus D70 & 9.0 & 549 & 0/3 & 250/0 & 8.3 & 549 & 2 & 2 & 0.10 & 15.4 \\
qugan\sus n395\sus D30 & 48 & \texttt{-|-} & 0/6 & \texttt{-|-} & 133 & \texttt{-|-} & 17 & \texttt{-|-} & 0.23 & \texttt{-|-} \\
qugan\sus n395\sus D50 & 33 & \texttt{-|-} & 0/6 & \texttt{-|-} & 74 & \texttt{-|-} & 10 & \texttt{-|-} & 0.24 & \texttt{-|-} \\
qugan\sus n395\sus D70 & 26 & \texttt{-|-} & 0/6 & \texttt{-|-} & 49 & \texttt{-|-} & 7 & \texttt{-|-} & 0.23 & \texttt{-|-} \\
swap\sus test\sus n83\sus D30 & 6.6 & 6.6 & 0/2 & 0/2 & 5.5 & 5.5 & 3 & 3 & 0.02 & 0.39 \\
swap\sus test\sus n83\sus D50 & 3.5 & 3.5 & 0/1 & 0/1 & 2.8 & 2.8 & 2 & 2 & 0.02 & 0.05 \\
swap\sus test\sus n83\sus D70 & 3.5 & 3.5 & 0/1 & 0/1 & 2.8 & 2.8 & 2 & 2 & 0.02 & 0.05 \\
swap\sus test\sus n115\sus D30 & 8.5 & 381 & 0/2 & 172/0 & 11 & 377 & 5 & 44 & 0.02 & 6.66 \\
swap\sus test\sus n115\sus D50 & 6.6 & 6.6 & 0/2 & 0/2 & 5.5 & 5.5 & 3 & 3 & 0.02 & 1.26 \\
swap\sus test\sus n115\sus D70 & 3.5 & 3.5 & 0/1 & 0/1 & 2.8 & 2.8 & 2 & 2 & 0.02 & 0.07 \\
swap\sus test\sus n361\sus D30 & 16 & \texttt{-|-} & 0/2 & \texttt{-|-} & 38 & \texttt{-|-} & 15 & \texttt{-|-} & 0.09 & \texttt{-|-} \\
swap\sus test\sus n361\sus D50 & 11 & 1376 & 0/2 & 624/0 & 19 & 1371 & 8 & 157 & 0.07 & 28.2 \\
swap\sus test\sus n361\sus D70 & 9.4 & 1288 & 0/2 & 584/0 & 13 & 1283 & 6 & 147 & 0.08 & 31.6 \\
wstate\sus n76\sus D30 & 9.9 & 9.9 & 4/0 & 4/0 & 8.8 & 8.8 & 3 & 3 & 0.01 & 14.5 \\
wstate\sus n76\sus D50 & 5.1 & 5.1 & 2/0 & 2/0 & 4.4 & 4.4 & 2 & 2 & 0.01 & 0.34 \\
wstate\sus n76\sus D70 & 5.1 & 5.1 & 2/0 & 2/0 & 4.4 & 4.4 & 2 & 2 & 0.01 & 0.55 \\
wstate\sus n118\sus D30 & 12 & 10 & 4/0 & 4/0 & 17 & 13 & 5 & 4 & 0.01 & 21.0 \\
wstate\sus n118\sus D50 & 9.9 & 9.9 & 4/0 & 4/0 & 8.8 & 8.8 & 3 & 3 & 0.01 & 31.2 \\
wstate\sus n118\sus D70 & 5.1 & 5.1 & 2/0 & 2/0 & 4.4 & 4.4 & 2 & 2 & 0.01 & 0.67 \\
wstate\sus n380\sus D30 & 28 & 19 & 4/0 & 4/0 & 96 & 52 & 23 & 13 & 0.04 & 20.8 \\
wstate\sus n380\sus D50 & 15 & 14 & 4/0 & 4/0 & 35 & 30 & 9 & 8 & 0.04 & 36.7 \\
wstate\sus n380\sus D70 & 13 & 13 & 4/0 & 4/0 & 21 & 21 & 6 & 6 & 0.04 & 50.2 \\
\end{longtable}
}

\section{Conclusion and discussion}\label{sec:conclusion}
We have introduced a novel method for identifying cut locations for quantum circuit cutting, with a primary focus on partitioning large circuits into three or more parts.
Under the assumption that the classical postprocessing function is decomposable, we have first derived the upper bound on the sampling overhead.
We have shown that this new bound significantly improves upon the previously known bound, which is based on the total number of cuts, by orders of magnitude in cases of three or more partitions.
The validity of our bound has been confirmed through the numerical simulations on circuits partitioned into three and four parts.

Next, based on the newly derived upper bound on the sampling overhead, we have defined a new objective function and proposed a method to identify cutting locations that minimize it. This objective function more appropriately reflects the sampling overhead in solutions of three or more partitions, compared to the total number of cuts used in the previous methods.
To validate our method, we applied it to 83 quantum circuits from the large scale category of QASMBench. We used the results obtained from Qiskit-addon-cutting (QAC) as the baseline.
We have observed that our method significantly outperforms QAC in terms of computation time by achieving a speedup ranging from $\sim 2 \times$ to $\sim 3080 \times$ depending on the circuit with an average speedup of $\sim 610 \times$.
In terms of solution quality with respect to $L_Q^{}$, our method has performed comparable to or better than QAC in all but 6 out of the 83 circuits.

In our derivation of  the upper bound on the sampling overhead, we assume independent cutting approaches~\cite{Peng_2020,Mitarai_2021,mitarai2021overhead}. Accordingly, the baseline for comparison (eq.~\ref{eq:overhead_K}) is also based on this assumption.
It has been shown that the sampling overhead can be reduced using techniques such as joint cutting and parallel cutting; see e.g. refs.~\cite{Lowe:2022lom, Piveteau_2024,harada2024doubly, Ufrecht_2023}.
Since these studies focus on bipartitions, extending joint cutting and/or parallel cutting to cases of more than two partitions may be a promising direction for future research.

\section*{Acknowledgments}
The views expressed in this research are those of the authors and do not necessarily reflect the official policy or position of PwC Consulting LLC.
We wish to thank Mitsuhiro Matsumoto, Tsuyoshi Kitano, Hana Ebi, Hideaki Kawaguchi, Hiroki Kuji, Shigetora Miyashita, Shin Nishio and Takaharu Yoshida for stimulating discussion and support. This work is supported by New Energy and Industrial Technology Development Organization (NEDO).

\bibliography{main}
\bibliographystyle{unsrt}

\appendix
\section{Derivation of sampling overhead for three or more partitions}\label{sec:appendix}

In this appendix we show eq.~\ref{eq:N_tot} by following the approach presented in ref.~\cite{Ufrecht_2023}.

Substituting eq.~\ref{eq:decompositions} into eq.~\ref{eq:exp_value} results in
\begin{align}
  \langle O \rangle =  \sum_{i_1^{}, \ldots, i_K^{}} a_1^{}(i_1^{}) \cdot \ldots \cdot a_K^{}(i_K^{}) \sum_s f(s) p(s | i_1^{}, \ldots, i_K^{}),\label{eq:exp_value_2}
\end{align}
where $p(s | i_1^{}, \ldots, i_K^{})$ is the probability of obtaining $s$ provided the circuits specified by $(i_1^{}, \ldots, i_K^{})$ are run,
\begin{align}
  p(s | i_1^{}, \ldots, i_K^{}) = \mathrm{tr}\bigl(P_s^{} \mathcal{W} \circ \mathcal{F}_K^{}(i_K^{}) \circ \ldots \circ \mathcal{F}_1^{}(i_1^{})    \circ \mathcal{U}(\rho)\bigr).
\end{align}
When the postprocess function is decomposed as eq.~\ref{eq:f_decompose}, the observable in eq.~\ref{eq:observable} can be manipulated as
\begin{align}
  O_c^{} \coloneq \sum_{s_c^{}} f_c^{}(s_c^{}) P_{s_c^{}}^{}, \ \ O = \prod_{c=1}^{R} O_c^{}. \label{eq:observable_2}
\end{align}
Moreover, the expectation value in eq.~\ref{eq:exp_value_2} can be written as
\begin{align}
  \langle O \rangle =  \sum_{i_1^{}, \ldots, i_K^{}} a_1^{}(i_1^{}) \cdot \ldots \cdot a_K^{}(i_K^{})  \prod_{c=1}^{R} \sum_{s_c^{}}f_c^{}(s_c^{}) p(s_1^{} \cdot \ldots \cdot s_R^{}  | i_1^{}, \ldots, i_K^{}).\label{eq:exp_value_3}
\end{align}
To make the manipulation simpler, we will consider a specific case of $R=3$ and $K=4$ as illustrated in Figure~\ref{fig:3partition}.
In this case, eq.~\ref{eq:exp_value_3} becomes
\begin{align}
  \langle O \rangle =  \sum_{i_1^{}, \ldots, i_4^{}} a_1^{}(i_1^{}) \cdot \ldots \cdot a_4^{}(i_4^{})  \prod_{c=1}^{R=3} \sum_{s_c^{}}f_c^{}(s_c^{}) p(s_1^{} | i_1^{}, i_2^{}, i_4^{}) p(s_2^{} | i_1^{}, i_2^{}, i_3^{}) p(s_3^{} | i_3^{}, i_4^{}).\label{eq:exp_value_4}
\end{align}
Let us define a random variable $F$ as
\begin{align}
  F \coloneq  \sum_{i_1^{}, \ldots, i_4^{}} a_1^{}(i_1^{}) a_2^{}(i_2^{}) a_3^{}(i_3^{}) a_4^{}(i_4^{})  F_1^{}(i_1^{}, i_2^{}, i_4^{}) F_2^{}(i_1^{}, i_2^{}, i_3^{}) F_3^{}(i_3^{}, i_4^{}),\label{eq:rand_1}
\end{align}
with the following random variables,
\begin{subequations}
\begin{align}
  F_1^{}(i_1^{}, i_2^{}, i_4^{}) & \coloneq  f_1^{}(S_1^{})|_{i_1^{}, i_2^{}, i_4^{}}^{} \label{eq:rand_2}, \\
  F_2^{}(i_1^{}, i_2^{}, i_3^{}) & \coloneq  f_2^{}(S_2^{})|_{i_1^{}, i_2^{}, i_3^{}}^{} \label{eq:rand_3}, \\
  F_3^{}(i_3^{}, i_4^{}) & \coloneq  f_3^{}(S_3^{})|_{i_3^{}, i_4^{}}^{} \label{eq:rand_4}.
\end{align}
\end{subequations}
Since these random variables can be independently sampled as
\begin{subequations}
\begin{align}
   \mathbb{E}(F_1^{}(i_1^{}, i_2^{}, i_4^{})) & \coloneq \sum_{s_1^{}}f_1^{}(s_1^{}) p(s_1^{} | i_1^{}, i_2^{}, i_4^{}) \label{eq:ef1}, \\
   \mathbb{E}(F_2^{}(i_1^{}, i_2^{}, i_3^{})) & \coloneq \sum_{s_2^{}}f_2^{}(s_2^{}) p(s_2^{} | i_1^{}, i_2^{}, i_3^{}) \label{eq:ef2}, \\
   \mathbb{E}(F_3^{}(i_3^{}, i_4^{}) & \coloneq  \sum_{s_3^{}}f_3^{}(s_3^{}) p(s_3^{} | i_3^{}, i_4^{}) \label{eq:ef3},
\end{align}
\end{subequations}
we find
\begin{align}
  \mathbb{E}(F) & \coloneq  \sum_{i_1^{}, \ldots, i_4^{}} a_1^{}(i_1^{}) a_2^{}(i_2^{}) a_3^{}(i_3^{}) a_4^{}(i_4^{})  \mathbb{E}(F_1^{}(i_1^{}, i_2^{}, i_4^{})) \mathbb{E}(F_2^{}(i_1^{}, i_2^{}, i_3^{})) \mathbb{E}(F_3^{}(i_3^{}, i_4^{})),\label{eq:rand_5} \\
 & = \langle O \rangle, \label{eq:rand_6}
\end{align}
where eq.~\ref{eq:exp_value_4} is referred.
Below we show that the standard deviation of $F$ can be bounded by $\epsilon$, namely
\begin{align}
  \mathrm{Var}(F) \le \epsilon^2_{},
\end{align}
if the number of sampling for each partition satisfies
\begin{subequations}\label{eq:N}
\begin{align}
  N_1^{} & \ge 3 \frac{(\kappa_1^{} \kappa_2^{} \kappa_4^{} )^2_{} \tau_3^{}}{\epsilon^2_{}}, \label{eq:n1}\\
   N_2^{} & \ge 3 \frac{(\kappa_1^{} \kappa_2^{} \kappa_3^{} )^2_{} \tau_4^{}}{\epsilon^2_{}}, \label{eq:n2}\\
    N_3^{} & \ge 3 \frac{(\kappa_3^{} \kappa_4^{} )^2_{} \tau_1^{} \tau_2^{}}{\epsilon^2_{}}. \label{eq:n3}
\end{align}
\end{subequations}
Within the partition $c=1$, the number of samples $n_1^{}(i_1^{}, i_2^{}, i_4^{})$ will be used to run the circuit specified by  $(i_1^{}, i_2^{}, i_4^{})$,
\begin{subequations}\label{eq:n123}
\begin{align}
  n_1^{}(i_1^{}, i_2^{}, i_4^{}) = \frac{ |a_1^{}(i_1^{})| |a_2^{}(i_2^{})| |a_4^{}(i_4^{})|}{\kappa_1^{} \kappa_2^{} \kappa_4^{}}N_1^{},\ \ \ \sum_{i_1^{}, i_2^{}, i_4^{}} n_1^{}(i_1^{}, i_2^{}, i_4^{}) = N_1^{}.
\end{align}
Similarly, for partitions $c=2$ and $c=3$,
\begin{align}
  n_2^{}(i_1^{}, i_2^{}, i_3^{}) & = \frac{ |a_1^{}(i_1^{})| |a_2^{}(i_2^{})| |a_3^{}(i_3^{})|}{\kappa_1^{} \kappa_2^{} \kappa_3^{}}N_2^{},\ \ \ \sum_{i_1^{}, i_2^{}, i_3^{}} n_2^{}(i_1^{}, i_2^{}, i_3^{}) = N_2^{}, \\
  n_3^{}(i_3^{}, i_4^{}) & = \frac{ |a_3^{}(i_3^{})| |a_4^{}(i_4^{})|}{\kappa_3^{} \kappa_4^{}}N_3^{},\ \ \ \sum_{i_3^{}, i_4^{}} n_3^{}(i_3^{}, i_4^{}) = N_3^{}.
\end{align}
\end{subequations}
The variance of $F_1^{}(i_1^{}, i_2^{}, i_4^{}) F_2^{}(i_1^{}, i_2^{}, i_3^{}) F_3^{}(i_3^{}, i_4^{})$, which is an each term in $F$ of eq.~\ref{eq:rand_1}, can be calculated as
\begin{align}
  \mathrm{Var}(F_1^{}(i_1^{}, i_2^{}, i_4^{}) F_2^{}(i_1^{}, i_2^{}, i_3^{}) F_3^{}(i_3^{}, i_4^{})) & \simeq \mathrm{Var}(F_1^{}(i_1^{}, i_2^{}, i_4^{})) +  \mathrm{Var}(F_2^{}(i_1^{}, i_2^{}, i_3^{}))  +\mathrm{Var}(F_3^{}(i_3^{}, i_4^{})) \label{eq:var_each_1}\\
  & \le \frac{1}{n_1^{}(i_1^{}, i_2^{}, i_4^{})} + \frac{1}{n_2^{}(i_1^{}, i_2^{}, i_3^{})} + \frac{1}{n_3^{}(i_3^{}, i_4^{})},\label{eq:var_each_2}
\end{align}
where it is assumed that each variance $\mathrm{Var}(F_i^{})$ is sufficiently controlled as  $\mathrm{Var}(F_i^{}) \ll 1$ in eq.~\ref{eq:var_each_1}. It is straightforward to show that
\begin{align}
  \mathrm{Var}(F) & = \sum_{i_1^{}, \ldots, i_4^{}} (a_1^{}(i_1^{}))^2_{} (a_2^{}(i_2^{}))^2_{} (a_3^{}(i_3^{}))^2_{} (a_4^{}(i_4^{}))^2_{}\mathrm{Var}(F_1^{}(i_1^{}, i_2^{}, i_4^{}) F_2^{}(i_1^{}, i_2^{}, i_3^{}) F_3^{}(i_3^{}, i_4^{})) \\
  & \le  \sum_{i_1^{}, \ldots, i_4^{}} (a_1^{}(i_1^{}))^2_{} (a_2^{}(i_2^{}))^2_{} (a_3^{}(i_3^{}))^2_{} (a_4^{}(i_4^{}))^2_{} \biggl( \frac{1}{n_1^{}(i_1^{}, i_2^{}, i_4^{})} + \frac{1}{n_2^{}(i_1^{}, i_2^{}, i_3^{})} + \frac{1}{n_3^{}(i_3^{}, i_4^{})} \biggr) \\
  & \le \epsilon^2_{},
\end{align}
where eqs.~\ref{eq:N}, \ref{eq:n123} are used.
Generalizing the above argument to any $R$ and $K$ is straightforward. In particular, eq.~\ref{eq:N} can be generalized to eq.~\ref{eq:N_tot}.

Finally, we compare with the result presented in ref.~\cite{Peng_2020}.
Ref.~\cite{Peng_2020} shows that the sampling overhead  for three or more partitions via time-like cutting under the assumption of eq.~\ref{eq:f_decompose} is upper-bounded by
\begin{align}
 N_{\mathrm{tot}}^{} \ge \frac{2(e-1)^2_{}(R 8^{d^{\prime}_{}}_{})^3_{} \ln{(6R 8^{d^{\prime}_{}}_{}})}{\epsilon^2_{}},
\end{align}
where $d^{\prime}_{}$ is the maximum number of time-like cuts on any partition. In the case of our example shown in Figure~\ref{fig:3partition}, this formula gives
\begin{align}
 N_{\mathrm{tot}}^{} \ge \frac{2.0 \times 10^{11}_{}}{\epsilon^2_{}},
\end{align}
which is much larger than our estimation in eq.~\ref{eq:Ntot_our}. In ref.~\cite{Peng_2020}, the upper bound is derived using the Hoeffding's inequality, whereas our approach differs in that we obtain it by computing the variance.
Moreover, while the result in ref.~\cite{Peng_2020} is based on the partition that requires the largest number of samples, our method provides the upper bound for each partition individually. Therefore, our result can be considered as an improvement on the upper bound proposed in  ref.~\cite{Peng_2020}.

\end{document}